\documentstyle[twocolumn,aps,prb]{revtex}
\draft
\begin{document}
\begin{title}
{Scaling  and finite-size-scaling in the two dimensional random-coupling
Ising ferromagnet}
\end{title}
\author{Jae-Kwon Kim}
\address
{School of Physics, Korea Institute for Advanced Study, \\
207-43 Cheongryangri-dong, Dongdaemun-gu, Seoul 130-012, Korea}
\maketitle

\begin{abstract}
It is shown by Monte Carlo method that 
the finite size scaling (FSS) holds  
in the two dimensional 
random-coupled Ising ferromagnet. 
It is also demonstrated that the 
form of universal FSS 
function constructed via novel FSS scheme depends 
on the strength of the random coupling for strongly 
disordered cases.
Monte Carlo measurements of thermodynamic 
(infinite volume limit) data of the correlation length
($\xi$) up to $\xi \simeq 200$ 
along with measurements of 
the fourth order cumulant ratio (Binder's ratio) at criticality 
are reported and analyzed in view of two 
competing scenarios. 
It is demonstrated that the data are almost exclusively
consistent with the scenario of weak universality.
\end{abstract}
\pacs{{\bf PACS numbers}: 75.40.Mg, 75.10.Nr, 05.50.Jk}

\section{Introduction}
	Random disorder plays important role in many interesting 
	phenomena of condensed matter physics. Some examples are
	Kondo problem and  metal insulator transition driven by random disorder.
	The two-dimensional (2D)
	randomly disordered Ising ferromagnet is 
	the simplest nontrivial statistical model including
	the effect of another type of fluctuation in addition to
	usual thermal fluctuation.
	By the random disorder is  meant either a random site
	dilution  or random-valued positive coupling. 
        The effect of
	the combined fluctuations of the thermal and
        quenched random disorder on the critical behavior of 
	the system has been the main subject of the studies.
	The two dimensional
        random coupling (or random bond) Ising ferromagnet 
	is defined by the Hamiltonian
	\begin{equation}
	H=- \sum_{<ij>} J_{ij} S_{i}S_{j},~~~J_{ij}>0,~~S_{i}=\pm 1,
	\end{equation}
	where the sum is over all the links of a lattice.
	To be more specific, for every link on a given lattice 
        one assigns randomly to $J_{ij}$ one of two different 
        ferromagnetic coupling constants --
        one that of the pure system $J$ and the other perturbed
	coupling $J^{\prime}$. In this way single realization of random
        distribution of coupling is made.
        Due to its random nature, obviously, the realization of the 
        disorder is uncorrelated.
	We are interested in the thermodynamic behavior  averaged
	over infinitely many different realizations of the 
        distribution of the coupling constants.
        
	McCoy and Wu\cite{MCO} made the first successful attempt
	for such a model, and
	proved that the specific heat ($C_{v}$) is non-divergent in the 2D
	Ising system with one-directional and correlated random
	bond disorder.
	According to McCoy and Wu, the divergence of $C_{v}$
	is caused by the coherence of all the bonds acting together,
	so destroying the coherence by introducing one or two
	directional bond disorders makes $C_{v}$ finite.

	Many authors\cite{DOT,SHA}, however, regarded the non-diverging 
	behavior of $C_{v}$ in the McCoy-Wu model
	as a characteristic of the one-dimensional correlated disorder.
	Especially, Shalaev, Shankar, and Ludwig (SSL)\cite{SHA},
	based on their observation that
	the continuum limit of the 2D random bond disordered Ising model
	is a certain type of Gross-Neveu model were able to obtain
        analytical prediction on the critical behavior of the disordered system.
        In this theory the disorder is regarded as a weak perturbation to
        non-interacting free fermion system (that is  equivalent to the
        2D pure Ising system), so that the theoretical prediction is supposed 
        to hold for weak random disorder. 
	The main theoretical prediction can be summerized as:
	\begin{eqnarray}
	\xi &\sim & t^{-\nu} |\ln t|^{\tilde{\nu}},
	~~\nu=1,~ \tilde{\nu}=1/2 \label{eq:ssl_cor} \\
	C_{v} &\sim & t^{-\alpha} |\ln|\ln~t||, ~~\alpha=0. \label{eq:spht} \\
	\chi &\sim & \xi^{2-\eta},~~\eta=1/4,  \label{eq:ssl_suc}
	\end{eqnarray}
	\label{eq:sp}
	where $t$ is the reduced temperature and $\chi$ is
	thermodynamic magnetic susceptibility.

	Note that in the context of SSL's theory  the critical behavior 
        is modified by the disorder only logarithmically, 
        with no difference in the values of the critical exponents
	from those of the corresponding pure system. 
        In the presence of extremely weak disorder the system must be 
        almost identical to the pure system, and the corresponding critical
        behavior must be almost indistinguishable from
	that of the pure system.
	This means that the modification of the critical behavior 
        in this case must take place
        in extremely narrow scaling regime near criticality only, 
        with the remaining scaling regime exhibiting the critical
        behavior of the pure system. 
	The conventional belief is that 
        Eq.(\ref{eq:ssl_cor})-Eq.(\ref{eq:ssl_suc}) remains to be
	valid for strongly disordered cases
        ---only the range of scaling region where they hold becomes 
	broader with the disorder.
	Hence the theory generally implies the presence of crossover 
        from the critical behavior of the pure Ising system to the asymptotic
        behavior Eq.(\ref{eq:ssl_cor})--Eq.(\ref{eq:ssl_suc}).
 
	The expressions reflecting the presence of the crossover thus read
	\begin{eqnarray}
	\xi &\sim & t^{-\nu} [1 + C|\ln(t)|]^{\tilde{\nu}},
	~~\nu=1,~ \tilde{\nu}=1/2 \label{eq:ssl_ceff} \\
	C_{v} &\sim & t^{-\alpha} \ln|1 + C\ln(t)| + C^{\prime}, ~~
	\label{eq:sp_eff} \\
	\chi &\sim & \xi^{2-\eta},~~\eta=1/4  \label{eq:ssl_seff}
	\end{eqnarray}
        The coefficients $C$ and $C^{\prime}$ as a function of the strength
        of disorder cannot be determined theoretically. 
        Nevertheless, obviously,
        $C$ increases with the strength of disorder, 
        and for a given value of $C$
        Eq.(\ref{eq:ssl_ceff}) reduces to its asymptotic form only when
	$t$ is very small.
        
	There is a more general argument, originally due to
        Harris\cite{HAR}, and later elaborated by semi-rigorous 
        approach\cite{CHAY}, on the effect of
        random quenched disorder to critical behavior.
        The argument can be formulated in the context of the 
        renormalization group theory, and it was obtained that
        weak, uncorrelated random disorder is irrelevant if
	\begin{equation}
	 D\nu > 2,   \label{eq:harris}
	\end{equation}
	where $\nu$ is the critical exponent of the correlation 
        length of the pure system.
	In our case D=2, $\nu=1$, the random disorder has 
        been regarded to be marginal leading to a logarithmic 
        correction to power-law critical singularity, being
	consistent with the prediction of SSL.

        However, especially in 2D, a presence of marginal perturbation
        in certain case such as in 
        Kosterlitz-Thouless type phase transition
        leads to continuously varying critical exponents. 
        A rigorously known example of 
        continuously varying critical exponents of $\gamma$ and $\nu$,
        but with the ratio $\gamma/\nu$ remains unchanged from
	the value of the pure Ising ferromagnet,
        takes place in some subspace of the coupling constants 
        of the 2D Ashkin-Teller model. 
	In this subspace, the model is equivalent to
	a certain annealed random coupling Ising ferromagnet.
	Also, many authors\cite{TIMO,TAMA} 
	obtained qualitatively different
	expressions for the critical behavior of $C_v$ from 
	that given by SSL. According to these expressions, $C_v$
	is non-diverging at criticality as is the case in 
	the McCoy-Wu model.

	It is now numerically established\cite{WANG,QUE,PAT,TALA} 
	that the value of 
	$\eta$ does not depend on the strength of disorder
	in both 2D random-coupled and site diluted Ising ferromagnets.
	Another generic feature obtained  from various numerical studies 
	\cite{WANG,QUE,PAT} is that other critical exponents
	such as $\gamma$ and $\nu$ increase 
	with the strength of disorder at least effectively. 
        In most cases this effective increase
	was interpreted as the multiplicative logarithmic correction.
	Whether these effectively varying critical exponents
	are due to crossover effect as claimed 
        in Ref.(\onlinecite{WANG,REIS,IGLOI,BALL,ROD,SELKE_97}) or 
	are indication of genuine novel critical behavior 
	induced by random disorder
	as claimed in Ref.(\onlinecite{PAT,KUHN}) has been a 
	controversial issue.
	Numerically it is very difficult to distinguish a pure power-law
	criticality with small critical 
	exponent from a logarithmic singularity, unless
	thermodynamic data in extremely deep scaling regime
	are available.
	Previously  claimed evidence for the predictions of SSL
	in most cases were nothing but the consistency with the logarithmic
	correction. This claimed consistency cannot rule out pure power-law
	singularity without the multiplicative logarithmic correction.
	Especially it was  observed that the {\it thermodynamic} values of
	the specific heat develops a non-diverging peak at a temperature
	larger than $\beta_c$\cite{PAT} for very strongly disordered
	random site diluted Ising model. 
        This contrasts clearly with Eq.(\ref{eq:sp_eff}),
	but agrees with other theoretical prediction\cite{TIMO,TAMA}.

	In this paper we attempt to clarify the unresolved
	issue  of the 2D random bond disordered Ising ferromagnet 
        by numerically determining the 
	functional form of the universal FSS function  defined
        in the context of novel FSS scheme\cite{KIM_FSS,KIM_PRE}.
	As long as the FSS is valid, the form of the FSS function 
	is unique for a given universality class.  
        Accordingly, if the system is 
	indeed in the same universality class
	irrespective of the strength of disorder then the form of each finite
	size scaling function must be identical for different strength of
	random disorder.
	We will also demonstrate that the thermodynamic data of $\xi$
	in sufficiently deep scaling regime tend to scale steeper than
	the asymptotic expression of SSL, Eq.(\ref{eq:ssl_cor}).
	In the sections to follow we elaborate on the FSS
	method employed here, give a detailed description of our
	Monte Carlo simulation, report our results, and finally
	conclude with some discussions.
	
\section{Generalized Finite Size Scaling}
	The fundamental assumption of FSS theory\cite{PRIV} is that
	$A_{L}(t)$, the value of some physical variable  $A$ on a
	finite lattice of linear size $L$, can be expressed as
	\begin{equation}
	A_{L}(t)=L^{\rho/\nu} f_{A} (s(L,t)),~~
	s(L,t)\equiv L/\xi(t) \label{eq:fun1}
	\end{equation}
	for a thermodynamic quantity $A$ which has a power-law 
	critical singularity
	$A(t) \sim t^{-\rho}$ with the reduced
	temperature  $t\equiv (\beta_c-\beta)/\beta_c$.
	Eq.(\ref{eq:fun1}) is valid for the values
	of $L$ and $\xi(t)$ that are sufficiently large;
	otherwise, there should be corrections to FSS. 

	Notice that using the critical form for $\xi$,
	$\xi(t) \sim t^{-\nu}$, one can
	rewrite the scaling variable $s(L,t)$ as
	\begin{equation}
	s(L,t)=(A(t)/L^{\rho/\nu})^{-\nu/\rho},
	\end{equation}
	so that Eq.(\ref{eq:fun1}) may be rewritten as
	\begin{equation}
	A_{L}(t)=A(t) {\cal F}_{A}(s(L,t)),    \label{eq:fun2}
	\end{equation}
	where the relation between the scaling functions $f_{A}$  and
	${\cal F}_{A}$ is given by
	\begin{equation}
	{\cal F}_{A}(s)=s^{\rho/\nu} f_{A}(s). \label{eq:rel1}
	\end{equation}
	For $A=\xi$, Eq.(\ref{eq:fun2}) shows that $\xi_{L}(t)/L$ is
	just a function of $\xi(t)/L$ and vice versa, and this leads to
	the relation\cite{KIM_FSS}
	\begin{equation}
	A_{L}(t)=A(t){\cal{Q}}_{A}(x(L,t)), \label{eq:fun}
	\end{equation}
	where $x(L,t)\equiv \xi_{L}(t)/L$ is the ratio of the correlation
	length  on a finite lattice to the linear size of the 
        lattice, and ${\cal{Q}}_{A}(x)$ is given by
	\begin{equation}
	{\cal{Q}}_{A}(x)={\cal F}_{A}(f^{-1}_{\xi}(x)).    \label{eq:rel2}
	\end{equation}
	Using the same observation, it is trivial to obtain\cite{SOK1}
	another equivalent form,
	\begin{equation}
	A_{bL}(t)=A_{L}(t){\cal{G}}_{A}(x(L,t)),
	\end{equation}
	where $b$ is a scaling factor and ${\cal{G}}_{A}(x)$ is another
	scaling function.

	It is evident that given $f_{A}$ one can determine the other
	two scaling functions from Eqs.(\ref{eq:rel1}) and (\ref{eq:rel2}).
	It is well-known that the standard scaling function $f_{A}$ is
	universal, so that
	all the other scaling functions, ${\cal{F}}_{A}$,
	${\cal{Q}}_{A}$, and ${\cal{G}}_{A}$ should be universal as well.
	It has also been argued\cite{HIL} that a certain asymptotic form
	of $f_{A}(s)$ can be expressed in terms of the critical exponent
	 $\delta$; by
	fitting this functional form one can extract an estimate for
	the critical exponent.

	It is worth stressing that use of the scaling function
	$\cal{Q}$ rather than $\cal{F}$ or $f_{A}$
        would be more convenient in many cases,
	particularly because one does not need the thermodynamic 
	correlation length
	to define the former.  
	Note that there is no explicit $t$ dependence of
	the scaling variables, 
	so that knowledge of 
	the critical temperature and $\nu$ is not
	required, and that $x$ becomes independent of $L$ at criticality.
	This $L$ independent value of $x$ at criticality, $x_{c}$, which
	characterizes a universality class for a given geometry,
	forms the upper bound of $x$.
	In other words, the scaling function
	$\cal{Q}$ is defined only over $0 \le x \le x_{c}$.
	A priori, the two
	limits of the scaling function $\cal{Q}$  are known for
	a continuous phase transition:
	$\lim_{x \to 0} {\cal Q}(x) \to 1$ and
	$\lim_{x \to x_{c}} {\cal Q} \to 0$, because $A_{L}$ converges
	to its thermodynamic value in the former case while $A(t)$ diverges
	in the latter case with $A_{L}(t)$ finite.
	${\cal{Q}}_{A}(x)$ turns out to be a monotonically 
	decreasing function of $x$
	for $A = \chi$ and $\xi$.

	It is important to realize that the knowledge
	of the scaling function
	${\cal Q}$ near $x \simeq 0$ plays as relevant
	role as that near
	$x \simeq x_{c}$ to the extraction of necessary information
	of the critical behavior in (deep) scaling region.
	It can be easily seen by noting that
	$x(L,t)$ for a fixed temperature arbitrarily close to
	criticality can be made arbitrarily close to zero by
	simply choosing a value of $L$
	sufficiently large.

	Eqs.(\ref{eq:fun2}) and (\ref{eq:fun}) do not include
	any critical exponents,
	so that one might conjecture that their validity can be
	extended to non-power-law singularities.
	Although a general proof of this
	conjecture is missing, for the two-dimensional (2D)
	$O(N)$ ($N >2$) spin model which exhibits an exponential
	critical singularity,
	L\"{u}scher\cite{LUS} obtained an
	explicit expression for the inverse correlation length (mass gap),
	\begin{equation}
	\xi_{L}(t) \sim \xi_{\infty}(t) \left[{e^{-4\pi F(x)}/x^{2}}
	\times \left(1+ {\cal{O}}(\log{L}/L^{2}) \right) \right]^{1/2} 
	\label{eq:lus}
	\end{equation}
	Eq.(\ref{eq:lus}) is reduced to Eq.(\ref{eq:fun})
	for the sufficiently large values of L as
	for  ${\cal{O}}(\log{L}/L^{2})$ to be safely neglected. 
	Note that 
	${\cal{O}}(\log{L}/L^{2})$ is comparable to statistical errors
	of typical Monte Carlo simulations at L=20 already.
	It is thus postulated\cite{KIM_FSS,KIM_PRE} that  
        Eq.(\ref{eq:fun}) is a generalized
	FSS form that is valid irrespective of the functional form
	of critical singularity,

	Many novel applications of FSS were obtained 
	thanks to Eq.(\ref{eq:fun}). The key observation is that 
	the scaling variable $x$ has no explicit 
	temperature dependence.
	First, it can be used to check the validity of
	FSS itself for a given physical system:
	one numerically calculates the scaling function
	${\cal{Q}}_{\xi}(x)$ at two
	arbitrary temperatures close to criticality. If the two 
	scaling function thus calculated ``collapse" unto a single
	function then Eq.(\ref{eq:fun}) must be valid. The demonstration
	of the validity of FSS for the 2D and 3D pure Ising models 
	in this context can be found in Ref.(\onlinecite{KIM_PRE}).
	The numerical determination of the 
        ${\cal{Q}}_{\xi}$ were carried out 
	for a variety of other statistical models such as 2D 
	Heisenberg and  XY models\cite{KIM_FSS}.
	This check is much more unambiguous than the standard 
	data collapse method of FSS on the model\cite{WISE}
        since it does not involve any 
	adjustable parameters such as
	the critical temperature and the critical exponent of the 
	correlation length $\nu$.
	Second, ${\cal{Q}}_{\xi}$ can be used to extrapolate infinite volume 
	limit (thermodynamic) values of various physical variables as first
	demonstrated in Ref.(\onlinecite{KIM_FSS,SOK1}).
	
\section{The  Simulation}
	We consider binary distribution of $J_{ij}$.
	Namely the value of $J_{ij}$  at a link $<ij>$ is randomly
	distributed between two positive values $J$ and $J^{\prime}$
	with probability $p$ and $1-p$ respectively.
	For $p=1/2$, the system is self-dual\cite{FISC}
	with the self-dual point given by
	\begin{equation}
	\tanh (J)= \exp (-2J^{\prime})
	\end{equation}
	A self-dual point equals the critical point of
	a system, provided the system has only one critical point.
	We fix $J=1$ and $p=1/2$ without any loss of generality,
	and consider three different values of $J^{\prime}$, i.e.,
	$J^{\prime}= 0.9$, 0.25, and 0.1.
	Accordingly the self-dual points (critical points) are
	given by $\beta_{c}=0.4642819\ldots$,
	$0.80705185\ldots$, and $1.10389523\ldots$  for
	$J^{\prime}=0.9$, 0.25, and 0.1 respectively.

	Our raw-data for each $J^{\prime}$ are obtained by choosing
	a realization of random $J^{\prime}$, then running
	Monte Carlo simulations in single cluster algorithm\cite{WOL} 
	with periodic boundary conditions; for each realization,
	measurements were taken over 10 000 configurations
	each of which was separated by 2-15 single cluster updatings
	according to auto-correlation times.
	The procedure is then repeated for different
	realizations of $J^{\prime}$.
	The average over all the different realizations converges
	as the numbers of the random realization
	increase; basically this mean value of
	a physical quantity is something physically interesting.
	To achieve the necessary precision for our FSS scheme
	the numbers of the different realizations we used are
	approximately 20-40, 150-250, and 300-1000 
	for $J^{\prime}=0.9$, 0.25, and 0.1 respectively; yet, in general,
	the fluctuation among different realizations of the random
	disorder is more significant than the statistical error
	for a given realization. This was particularly the case
	for $J^{\prime}=0.1$. Nevertheless, 
        the average over different realizations obviously
        converges to a physically meaningful value.
	Our quoted error bars in our numerics are obtained by
	standard jack-knife method, taking into consideration
	the variation with different realization only.
	The smallest and largest values of L we used to extract 
	our thermodynamic values are 20 and 400 respectively. 
	For $J^{\prime}=0.05$ it turns out that data of very large 
	$\xi$ are not necessary, yielding compelling results
	with the data for $\xi \lesssim 50$ already. 
	Thus the FSS extrapolation method is not used for this $J^{\prime}$ 
	and 250-500 different realizations of distribution of $J^{\prime}$ 
	turns out to be sufficient.

	Determination of $A_{\infty}$ and the size dependence of 
	$A$ upon L, $A_{L}$, is essential to the computation of
	${\cal{Q}}_{A}$. The measurements of the correlation length
	must be taken into consideration for the computation
	of the scaling variable $x$. The correlation lengths
	are measured employing the standard formula of
	second moment correlation length\cite{KIM_PRE}.
        For the simplicity of
	our presentation, here we mainly focus on the 
	correlation length.
	For each $J^{\prime}$, we chose three different
	inverse temperatures 
	for the computations of the scaling
	variable and scaling function. 
	It is important to select  the values of the temperatures
	so that they are in the scaling regime.
	The criterion for this, which is known from the studies of well-known
	pure systems such as 2D Ising and three-state Potts models, 
        is that the thermodynamic value of the
	correlation length at a temperature in the scaling regime
        should be sufficiently large, namely
	$\xi \gtrsim 5$. 

\section{Result and Analysis}
	The ${\cal {Q}}_{\xi}(x)$ are calculated for $J^{\prime}=0.9,
	0.25$, and 0.1.
	For an illustration, the size dependence
	of the $\xi$ and $\chi$,
        and the results of the computations 
	of $x$, ${\cal {Q}}_{\xi}(x)$, and ${\cal {Q}}_{\chi}(x)$  
	obtained for $J^{\prime}=0.25$ and $\beta=0.77$
	and for $J^{\prime}=0.1$ and $\beta=1.04$ 
	are tabulated in Table(\ref{table_size}).
	It is worth noticing that both $\xi_L$ and $\chi_L$
	are monotonically increasing function of L. 
	(This is the case
	even in terms of the scaling variable $s$.)
	The $L$ dependence becomes weaker
	with increasing $L$, and becomes vanishingly small for 
	sufficiently large L. 
        This L independent value within the statistical 
	errors is the corresponding thermodynamic value.
	In Table(\ref{table_size}) it is shown that one needs
	$L/\xi_L \gtrsim 10$ for the accurate measurements of
	the thermodynamic values. The value of the ratio (L to $\xi_L$)
        is larger than in the corresponding pure system
	that is approximately 7\cite{KIM_PRE}. It is worth stressing that
	the value of the ratio beyond which a measurement 
	becomes thermodynamic is temperature independent 
	due to Eq.(\ref{eq:fun}).
	One thus needs to simulate at least with L larger than 
	10 times of the correlation length to make the data 
	be free of finite size effect.
	
	The plots of ${\cal {Q}}_{\xi}(x)$ 
	for the three values of $\beta$ in the scaling regime
	are shown in Fig.(\ref{fig:fss_25}) for $J^{\prime}=0.25$.
	Clearly each data set belonging to different
	$\beta$ collapses onto a single curve
	that is the universal FSS function for the value
	of $J^{\prime}$.
	Thus  the validity of the FSS is verified for 
	the $J^{\prime}$. We repeated similar procedure
	for the other two values of $J^{\prime}$ and observed
	similar data-collapse.
	For comparison, the FSS function for each $J^{\prime}$
	is shown in our Fig.(\ref{fig:fss_all}).  
	It is observed that the FSS function for $J^{\prime}=0.9$
	is indistinguishable from that of the corresponding pure system
	that was  calculated in Ref.(\onlinecite{KIM_PRE}).
	However, for stronger disorder the FSS scaling function clearly
        depends on $J^{\prime}$.
	The ansatz for our scaling function is
	\begin{equation}
	{\cal{Q}}_{\xi}(x)=1+c_{1}x+c_{2}x^2+c_3x^3+c_4x^4.
	\end{equation}
	The values of the coefficients $c_i (i=1,\ldots,4)$ 
	are calculated by fitting our data to the ansatz
	for each value of $J^{\prime}$. 

	The thermodynamic correlation lengths are measured
	directly (without using any extrapolation method)
        for the data roughly over the range $\xi \lesssim 40$. 
	With the knowledge of the scaling function available,
	the thermodynamic values 
	closer to criticality can now be easily 
	estimated by the use of the single-step
	FSS extrapolation method\cite{KIM_FSS}. 
	For the each value of $J^{\prime}$ the thermodynamic values
	of the correlation length are evaluated
	over the ranges
	$5.7(1) \le \xi \le 204(2)$ ([5.7(1), 204(2)]),
	[5.8(1), 217(3)], and [5.0(1), 203(5)].
	For the $J^{\prime}=0.05$ the thermodynamic data were obtained 
	by direct measurement under the thermodynamic condition 
	$L/\xi_L \ge 12$.
	which is over the range $5.52(2) \le \xi \le 47.64(29)$.
	Our thermodynamic data are summerized in 
	Table(\ref{table_thermo}).

	$\ln \xi(t)$ as a function of
	$|\ln t|$ is plotted in Fig.(\ref{fig:power}).  
	The slope of each straight line corresponds to the value of
	$\nu$. It is evident that $\nu$ increases with decreasing
	$J^{\prime}$ at least effectively.
	Fixing the critical points at the self-dual points in the
	$\chi^{2}$ fits and assuming a pure power-law type
	critical behavior,
	we obtain $\nu=1.01(1)$, 1.10(2), 1.20(3), 1.34(6) 
	for $J^{\prime}=0.9$, 0.25, 0.1, and 0.05 respectively.
	Assuming a scaling function with a nonconfluent
	correction term, e.g., $\xi(t) \sim t^{-\nu}(1+a t)$,
	yields the estimate of the critical
	exponent, e.g., $\nu=1.08(4)$ and $\nu=1.17(5)$ 
	for $J^{\prime}=0.25$ and $J^{\prime}=0.1$ respectively.
	Notice that for $J^{\prime}=0.9$ the estimated value of
	$\nu$ is virtually the same as that in the pure system.

	Although our thermodynamic data are extraordinarily broad
	compared to standard Monte Carlo or series expansion study
        can usually make available,
	it turns out that the data fit to a pure power-law as well
	as to the form with the multiplicative logarithmic correction.
	This can be easily seen by simple numerical experiment as well.
	Namely, generate some numbers according to a certain power-law
	with the value of $\nu$ slightly larger than 1,
	and  fit them  to a multiplicative 
	correction of the form Eq.(\ref{eq:ssl_ceff}).
	One can always find acceptable values of
	$C$ and $\tilde{\nu}$ over quite broad ranges, making it
	prohibitly difficult to get accurate
	estimate of the logarithmic exponent. 
	With the use of the asymptotic form,Eq.(\ref{eq:ssl_cor}),
        can one get much shaper estimate of $\tilde{\nu}$ than  with
	Eq.(\ref{eq:ssl_ceff}). The problem is that one does not
	know a priori whether all the data are beyond 
	the crossover point.
	Nevertheless, we find that use of the asymptotic form
	rather than Eq.(\ref{eq:ssl_ceff}) is quite useful 
	for our purpose.

	A useful observation is that $[1 + C|\ln(t)|]^{\tilde{\nu}}$ 
	($\tilde{\nu} > 0$) is less singular than $|\ln t|^{\tilde{\nu}}$ 
	for {\it any} range of $t >0$. Namely, note that
	\begin{equation}
	\left( 1 + C|\ln t| \right)^{\tilde{\nu}} \sim
	|\ln t |^{\tilde{\nu}} \left ( C + 1/ |\ln t| \right)^{\tilde{\nu}},
	\label{eq:obs}
	\end{equation}
	and that the second term on the right hand side of 
	Eq.(\ref{eq:obs})
	suppresses the singularity for any physical value of $t$. 
	Thus the asymptotic singularity Eq.(\ref{eq:ssl_cor})
        is always {\it more} singular than the mixture
	of the singularities, Eq.(\ref{eq:ssl_ceff}).
	Accordingly, if a thermodynamic correlation data
	turn out to be more singular than the asymptotic singularity
	{\it in an arbitrary portion} of the scaling regime, 
	then the the prediction of SSL must be invalidated.

        In Fig.(\ref{fig:ratio}) we plotted 
	$\xi(t)/(t^{-1} |\ln t|^{0.5}$)
        for $J^{\prime}=0.25$, 0.1, and  0.05.
	For the $J^{\prime}=0.25$, it is observed that the value of the 
        ratio decreases
	monotonically until temperature is very close to
	criticality, but starts to increase with further approaching
	to the criticality. 
	This is surprising in view of
	the SSL's picture, because the figure shows none of the data
	are either in the asymptotic regime or in the scaling regime
	of the pure system.
	For the $J^{\prime}=0.1$, we find that the data less close to 
	the criticality are consistent with the prediction of SSL, 
	but start to deviate from it as $t \to 0$.
	For the $J^{\prime}=0.05$, we observe that the data scale
	faster than the asymptotic form for all our data.
	We thus lead to the picture that seemingly consistency
	with the logarithmic correction for weak disorder 
	starts to become invalidated in sufficiently deep scaling regime.
	For the very strong disorder, any portion of data are inconsistent
	with the asymptotic form.

	Binder's cumulant ratio at criticality, denoted by $U^{(4)}_L$,
        is another universal quantity\cite{PRIV,BIN}.
	For each $J^{\prime}$ 
	we measured it at the critical point with varying $L$ 
	(Table\ref{table_binder}).
	It is observed that given $J^{\prime}$ $U^{(4)}_L$ is invariant 
	with $L$ within the statistical errors, and  that it tend to 
	increase uniformly with decreasing $J^{\prime}$. 
	The value for $J^{\prime}=0.9$
	is indistinguishable from the pure case, as is the case for
	the scaling function ${\cal{Q}}_\xi$.

\section{Conclusion and Discussion}
We have obtained clean numerical evidence that
FSS holds for quenched random disordered Ising ferromagnet.
The universal FSS function 
is found to be dependent upon the strength of 
disorder for strongly disordered cases.
For very weak disorder, however, it appears that disorder 
does not induce new universality class to the resolution of our data.
The behavior of Binder's cumulant ratio is in complete
agreement with the feature obtained from the FSS scaling function.
We also show that the singularity of $\xi$ is steeper than
the theoretical prediction for the data sufficiently close
to criticality, which almost certainly rules out
the validity of SSL in the strong disordered system.
Our result of varying exponent $\nu$ combined with the established 
fact of the invariance of $\gamma/\nu$ supports the scenario
of weak universality\cite{SUZ}. The same numerical evidence was
obtained for the random-coupling three-state Potts 
ferromagnet as well\cite{KIM_RBP}.	

As mentioned in the introduction most of previous numerical studies 
claiming for the evidence for the SSL came from the observation 
of the consistency of the logarithmic correction.
The consistency, however, cannot disprove the pure power-law
singularity that is essentially more steeper than the 
logarithmic correction. 
A very strong claim for the
evidence for the SSL was made in a recent high temperature
expansion study of the same model\cite{ROD}. 
What the authors of the paper observe is the monotonic 
increase of the effective value of $\gamma$ 
with the strength of disorder.
On the other hand, they claimed that the same data fitted 
to the prediction of SSL, Eq(\ref{eq:ssl_suc}), 
give rise to the value of the 
logarithmic exponent independent of the disorder.
This is unlikely to be mathematically correct because 
the effective increase of $\gamma$ leads to the 
effective increase of the logarithmic exponent, as
can be easily checked by simple numerical experiment\cite{COM}.

The study of the FSS behavior of the specific heat
\cite{WANG,INO,PLE,BALL,STAU}, 
instead of its thermodynamic behavior, 
should also be treated with caution.
In Ref.(\onlinecite{PAT})
it was shown that the size dependence of the specific heat is
different from that of the correlation length or the magnetic
susceptibility. Namely, the specific heat increases with 
$s \equiv L/\xi(t)$ for smaller $s$ but starts to decreases 
with it for larger $s$.
In view of FSS, Eq.(\ref{eq:fun2}),
what this implies is the existence of the upper bound value 
of the scaling variable $s$ below which the specific heat 
increases with L.
Thus, for any value of L, one is bound to observe monotonic 
increase of the specific heat at criticality where $\xi$ diverges 
(in other words, $s$ is smaller than the upper bound value for 
any finite L when $\xi \to \infty$), and this
has nothing to do with the actual divergence of the 
specific heat at criticality. 
From the study of the FSS behavior of the pure system\cite{KIM_PHYA}, 
we also note that 
FSS of the specific heat 
gives rise to less accurate result than 
that of $\xi$ or $\chi$.

\newpage
\begin{table}[t]
\caption{Size dependence of various physical quantities and 
	the computed scaling variable and scaling functions 
        at $\beta=0.77$ with $J^{\prime}=0.25$ (the upper part) and 
        at $\beta=1.04$ with $J^{\prime}=0.1$ (the lower part).
        The L independent value 
	(within the statistical errors) is the 
	corresponding thermodynamic value, which
        are $\xi=18.9(2)$ and $\xi=25.4(3)$ respectively 
	for $\beta=0.77$ and $1.04$.}
\label{table_size}
\begin{tabular} {cccccc}
$L$  &$\xi_L$  &$\chi_L$  &$x$  &${\cal{Q}}_{\xi}$  &${\cal{Q}}_{\chi}$\\ 
	\hline
20	&11.51(9)	&142.0(8)	 &0.576(5)   &0.609(11)   &0.311(5) \\
24	&12.69(9)       &178.0(1.2)	 &0.529(4)   &0.671(12)   &0.389(4) \\
30	&14.21(12)	&231.1(1.7)      &0.474(4)   &0.752(14)  &0.506(6)\\
34	&15.02(10)      &262.4(1.7)      &0.442(3)   &0.794(14)  &0.574(6)\\
40	&15.77(8)	&301.5(1.8)      &0.394(2)   &0.834(13)   &0.660(7)\\
50	&16.94(10)	&356.3(2.9)      &0.339(2)   &0.896(15)  &0.780(10)\\
60	&17.42(10)	&390.1(3.3)      &0.290(2)   &0.922(15)  &0.854(11)\\
70	&17.78(13)	&408.6(4.0)      &0.254(2)   &0.941(17)  &0.894(13)\\
80	&18.20(18)	&428.4(5.6)      &0.228(2)   &0.963(20)  &0.937(16)\\
150     &18.71(14)      &456.0(5.0)      &0.125(1)   &0.992(18)  &0.998(15)\\
200	&18.86(11)      &456.3(2.5)      &0.094(1)   &0.998(16)  &0.998(10) \\
250     &18.85(17)      &456.2(3.8)      &0.075(1)   &0.997(21)  &0.998(14) \\ \hline
20      &12.61(9)       &154.0(8)   &0.630(5)   &0.498(7)    &0.198(2) \\
30      &16.17(14)      &267.1(2.1) &0.539(5)   &0.639(10)   &0.344(5) \\
50      &20.41(23)      &467.7(5.8) &0.408(5)   &0.807(15)   &0.603(11) \\
80      &23.01(19)      &647.1(7.2) &0.288(2)   &0.909(15)   &0.834(15) \\
120     &24.16(17)      &720.1(7.5) &0.201(1)   &0.955(14)   &0.928(16) \\
160     &24.55(15)      &746.5(6.4) &0.153(1)   &0.970(14)   &0.962(14)\\
200     &24.86(14)      &759.0(5.9) &0.124(1)   &0.983(13)   &0.978(14)\\
240     &25.17(14)      &777.7(4.8) &0.105(1)   &0.995(13)   &1.002(13)\\
300     &25.35(16)      &776.2(4.5) &0.084(1)   &1.002(14)   &1.000(12)\\
360     &25.29(29)      &775.3(7.8) &0.070(1)   &1.000(19)   &0.999(16)
\end{tabular}
\end{table}

\onecolumn
\begin{table}
\caption{ Selection of thermodynamic $\xi$ 
	  for $J^{\prime}$=0.9, 0.25,  0.1 and 0.05. For the $J^{\prime}$=0.05
          all the results are obtained by direct measurements (without the use of the FSS
	  extrapolation method).}
\label{table_thermo}
\begin{tabular}{cc|cccccccc}
$J^{\prime}=0.9$ &$\beta$  &0.42  &0.44  &0.45  &0.455 &0.46 
			&0.462  &0.463   \\
		 &$\xi$  &5.76(5) &10.7(1)  &18.4(1) &28.3(2)
		   &61.9(3)  &116(1)  &204(2) \\ \hline		 
$J^{\prime}=0.25$ &$\beta$ &0.70  &0.74 &0.77 &0.78 
			   &0.79  &0.80 &0.803   \\
	          &$\xi$   & 5.80(5) &9.72(7) &18.9(1) &26.7(2) 
                      &43.6(2) &116(1) & 217(3) \\ \hline
$J^{\prime}=0.1$ &$\beta$ &0.87  &0.92  &1.0  &1.04 
			&1.06  &1.08   &1.093 \\
		 &$\xi$  &5.05(7)  & 6.95(8) &14.20(10)  &25.4(3)
		 &39.4(4) &80.5(1.4)  &203(5) \\ \hline
$J^{\prime}=0.05$ &$\beta$ &1.00  &1.10     &1.20       &1.25    &1.28  
					& -    & - \\
		  &$\xi$   &5.52(2)     &9.15(9)  &18.40(11)  &30.75(18) &47.64(29) 
					& -  & - \\
\end{tabular}
\end{table}

\twocolumn
\begin{table}
\caption{Binder's cumulant ratio at the self-dual points, for
         the three values of $J^{\prime}$.
         Note that $U_{L}(t=0)$ for each $J^{\prime}$ does not
         vary with L within the statistical errors,
         thus showing that each self-dual
         point is indeed the critical point. Also,
         it is clear that $U_{L}(t=0)$ increases with
         decreasing $J^{\prime}$,
         although for $J^{\prime}=0.9$ it is hardly distinguishable
         from the value of the pure system. For $J^{\prime}=0.25$ we extended
	 the measurements up to L=400, which does not show any sign of crossover.}
\label{table_binder}
\begin{tabular}{ccccc}
L    &$J^{\prime}=1.00$   &$J^{\prime}=0.90$    &$J^{\prime}=0.25$    &$J^{\prime}=0.10$ \\
        \hline
20  &1.8324(6)   &1.834(1)        &1.850(3)             &1.862(3)   \\
40  &1.8321(6)   &1.833(2)        &1.846(3)             &1.858(3)   \\
60  &1.8317(5)   &1.832(1)        &1.852(3)             &1.854(4)    \\
80  &1.8318(5)   &1.833(1)        &1.847(3)             &1.862(4)   \\
100 &1.8316(6)   &1.832(2)        &1.849(3)             &1.858(4)   \\
200 &  -         & -              &1.844(3)             & -         \\
400 &  -         & -              &1.845(2)             & -
\end{tabular}
\end{table}


\begin{figure}
\caption{Numerical calculation of ${\cal Q}_\xi$ for $J^{\prime}=0.25$,
         at three arbitrary inverse temperature $\beta$=0.74, 0.77,
         and 0.79 in the scaling regime. Our figure clearly indicates
	 that the scaling function does not have explicit temperature
	 dependence. For all the values of $\beta$, the smallest value
	 of $L$ used is 20.}
\label{fig:fss_25}
\end{figure}

\begin{figure}
\caption{The ${\cal Q}_{\xi}(x)$ for the three values of $J^{\prime}$.
	The dependence of the scaling function on $J^{\prime}$ is obvious.}
\label{fig:fss_all}
\end{figure}

\begin{figure}
\caption{$\ln\xi$ versus $|\ln t|$.
	The dotted lines represent the results of
	the best $\chi^{2}$ fits assuming
	pure power-law type singularity. The values of the slope
        , which correspond to the values of $\nu$, are 1.00, 1.10,
	1.20, and 1.34 respectively for $J^{\prime}=0.90$, 0.25, 0.10, and 0.05.}
\label{fig:power}
\end{figure}

\begin{figure}
\caption{The ratio $\xi(t)/(t^{-1} |\ln t|^{1/2})$ for 
	 $J^{\prime}=0.25$, 0.1, and 0.05.
	 The data for $J^{\prime}=0.1$ and 0.05 are uniformly shifted so that
	 the difference in the data points are clearly visible in a single figure.
	 Here increasing value of the ratio as $t$ becomes smaller
	 is an evidence that the data over the regime are inconsistent
	 with the prediction of SSL. The tendency of the increasing
	 ratio with $t$ becoming sufficiently
         small is observed for $J^{\prime}=0.25$ and 0.1, 
	 showing that the seemingly consistency with the logarithmic
	 correction of SSL for the larger values of $t$
         becomes invalidated in sufficiently deep scaling regime.  
	 For the $J^{\prime}=0.05$,
         all the data scales steeper than the asymptotic scaling, showing that
	 SSL cannot be correct for any data in 
         the scaling regime for this $J^{\prime}$.}
\label{fig:ratio}
\end{figure}

\begin{references}
\bibitem{COM} We generated numerical data from computer
              assuming pure power-law singularity $\chi \sim t^{-\gamma}$
              with three arbitrary $\gamma$=2.1, 2.3, and 2.5
	      (corresponding to $J^{2}/J^{1}\simeq$  5, 8 and 10 in their
	      paper\cite{ROD}), with the range of $t$ over $0.01 \le t \le 0.05$
	      and with relative errors one percent.
	      It is found that the data are fitted to the asymptotic 
	      form of SSL, Eq.\protect(\ref{eq:ssl_suc})
	      to yield the estimates of the logarithmic exponent 
              1.31, 2.06, and 2.44 respectively. 
	      The estimates of the logarithmic exponent depend on the 
	      choice of the range of $t$, being larger for $t$ smaller. 
              Nevertheless, the feature of its monotonic increase with
	      decreasing $J^{\prime}$ is invariant with the choice of
	      the range of $t$.
\end{references}
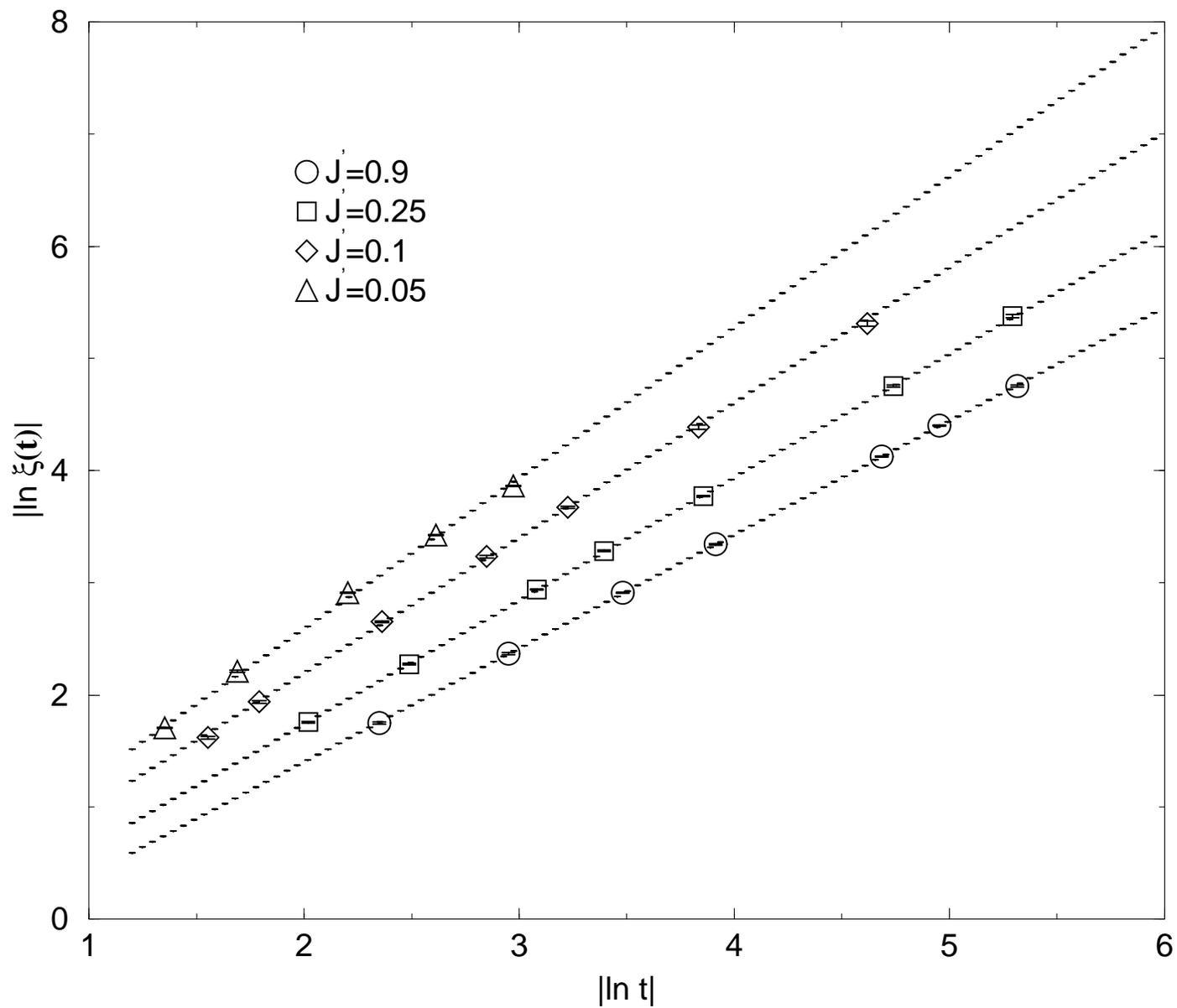
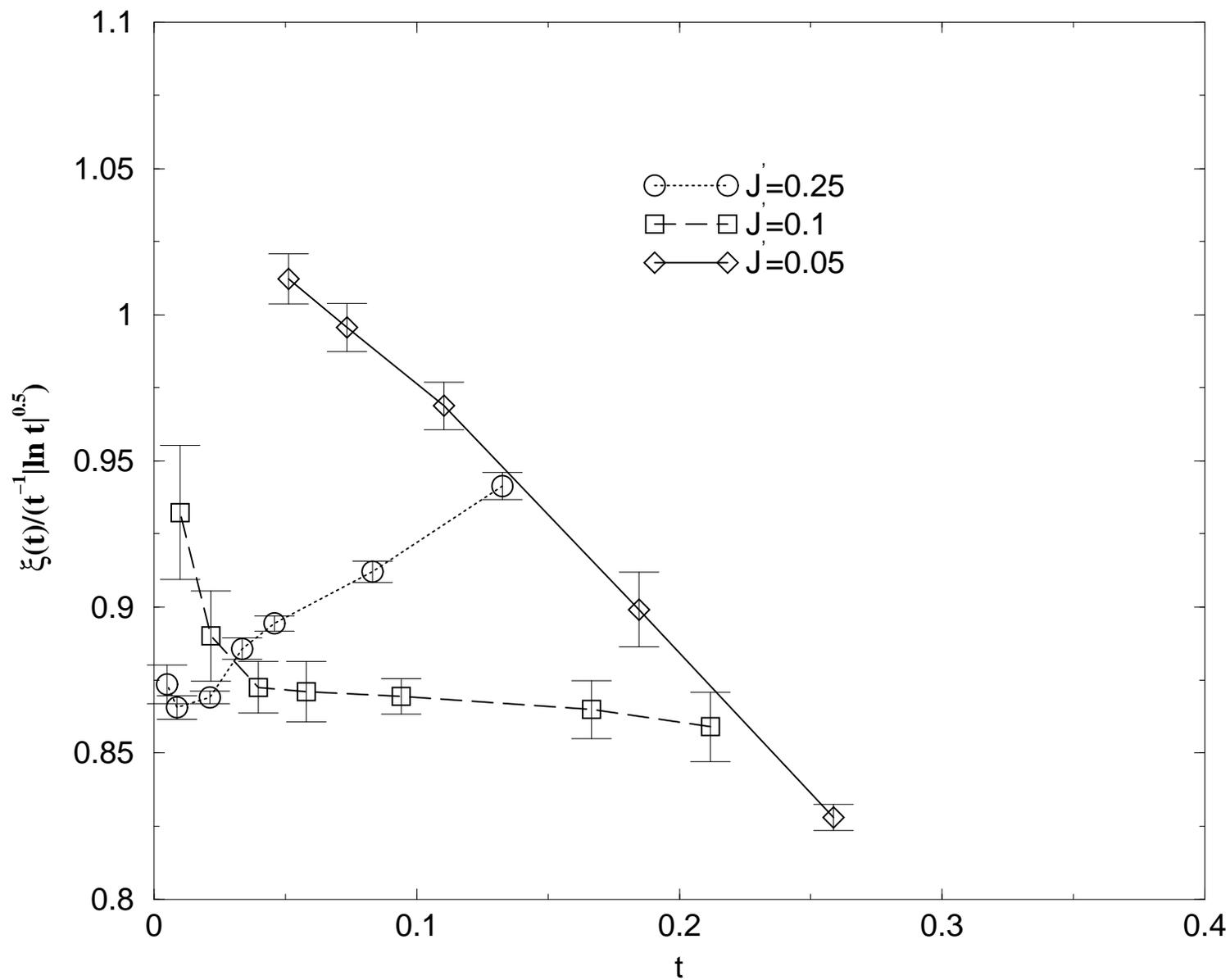
\end{document}